\documentclass[amsmath, aps,twocolumn,prl]{revtex4}
\usepackage{graphicx}
\usepackage{times}
\usepackage{epsfig}
\usepackage{multirow}
\usepackage{dcolumn}
\usepackage{booktabs}
\usepackage{revsymb}

\begin{document}

\title{Thermal corrections of lowest order for a helium atom}

\author{D. Solovyev}
\email{d.solovyev@spbu.ru}
\affiliation{Department of Physics, St.Petersburg State University, St.Petersburg, 198504, Russia}

 \author{T. Zalialiutdinov}
\affiliation{Department of Physics, St.Petersburg State University, St.Petersburg, 198504, Russia}
 \author{A. Anikin}
\affiliation{Department of Physics, St.Petersburg State University, St.Petersburg, 198504, Russia}

\begin{abstract}
In this paper the new type of thermal corrections for the helium and helium-like atomic systems are introduced. These are thermal one-photon exchange between the bound electrons and nucleus as well as between the bound electrons induced by the blackbody radiation (BBR).  All the derivations are given within the rigorous QED theory. It is shown that these thermal corrections are the same order in powers of $\alpha$ (fine structure constant) as the well-known BBR-induced Stark shift but the different behaviour in temperature. The numerical results presented in this paper make possible to expect their significance for modern experiments and testing the fundamental interactions in helium.
\end{abstract}
\maketitle

Helium is a few-electron atomic system which is the subject of present theoretical and experimental studies. The great success in experimental measurements of the transition frequencies in the helium atom \cite{Roo,Zheng} makes it necassery to carry out very precise theoretical calculations \cite{Drake_1999,Korobov_2000} and vice versa. At the present days, experimental accuracy reaches the level of several parts in $10^{-12}$ and requires a detailed analysis of various relativistic, quantum electrodynamics (QED) and etc corrections, see for example \cite{PYP-2016,PYP-2017}, which are usually evaluated in the framework of nonrelativistic theory. Progress in theoretical calculations and spectroscopic measurements of the energies of bound states in a helium atom makes it possible to determine accurately the charge radii of the nuclei of the ${}^3$He and ${}^4$He isotopes \cite{MWDrake,PK-Ch}.

The very precise theoretical calculations of the binding energies, the fine structure, and the isotope shift of the low-lying states of helium are inclined to serve as an independent tool for testing of fundamental interactions. Similar to well-studied one-electron atomic systems the measured transition frequencies should be compared with the theoretical calculations pursuing the search of possible discrepancy \cite{PPY-2017}. The quantum electrodynamics theory developed during last decades for the helium and helium-like ions allows the calculations of different corrections arising within expansion in powers of the fine structure constant $\alpha$ and the electron-to-nucleus mass ratio $m/M$ upto the effects of the order of $\alpha^6 m^2/M$ and $\alpha^7m$ \cite{PPY-2017}. 

Such extraordinary calculations however pay attention to the effects of the other type: corrections induced by the external blackbody radiation (BBR) field. The influence of the BBR field is well-known in atomic physics, it is manifested in the existence of a Stark shift of bound states. The theory and corresponding calculations for one- and few-electron atoms were presented in \cite{farley} in the framework of the Quantum Mechanical (QM) approach. These calculations were continued for the case of atomic clocks (many-particle systems) in \cite{Saf,SKC,Porsev} and are the subject of theoretical investigations in present days. Not long ago, the QED derivation of the Stark shift induced by the BBR field was performed in \cite{SLP-QED} and, subsequently, applied to calculations in the helium atom \cite{jphysb2017}. 

Recently, the QED theory of thermal energy shifts induced by the BBR field was presented in \cite{S-TQED} for the one-electron atomic systems. In particular it was found that the thermal interaction of a bound electron with a nucleus can lead to an energy shift exceeding the corresponding Stark shift. Although the theory developed in \cite{S-TQED} is a continuation of thermal quantum electrodynamics (TQED), see, for example, \cite{Dol,Don,DHR}, the effect of thermal interaction has not been previously considered. An experimental verification of this effect for a hydrogen atom was also proposed in \cite{S-TQED}. In this paper, the same corrections are investigated for two-electron atomic system. Increasing experimental accuracy \cite{Roo,Zheng} makes it possible to observe this effect in helium. It can be expected that thermal corrections of this type can serve for further testing of fundamental interactions on helium atom \cite{PPY-2017}.

\begin{figure}[hbtp]
	\centering
	\includegraphics[scale=0.15]{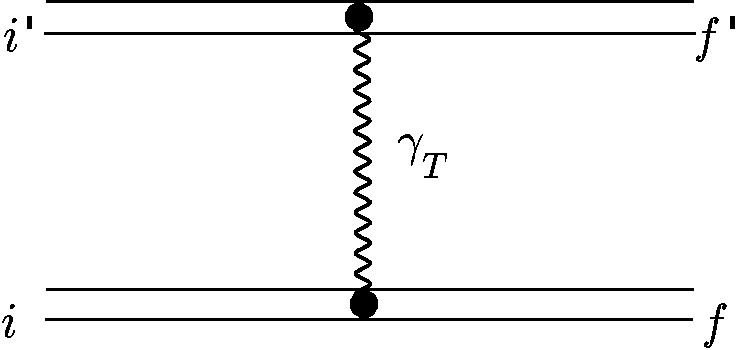}
	\caption{The Feynman diagram depicting the one thermal photon exchange between the bound electrons. The double solid lines denote the bound electron. The wavy line with the index $\gamma_T$ means the thermal photon. Indices $i(i')$ and $f(f')$ characterize the initial and final states of bound electrons, respectively.}\label{Fig1}
\end{figure}

We start by considering the interelectron interaction given as the one thermal photon exchange between two bound electrons. Such interaction can be presented schematically by the Feynman graph in Fig.~\ref{Fig1}.
The $S$-matrix element is
\begin{eqnarray}
\label{1}
S_{fi} = (-ie)^2\int d^4x_1 d^4x_2\left(\bar{\psi}_f(x_1)\gamma^\mu\psi_i(x_1)\right)\times
\\
\nonumber
 D_{\mu\nu}^\beta(x_1,x_2) \left(\bar{\psi}_{f'}(x_2)\gamma^\nu\psi_{i'}(x_2)\right),
\end{eqnarray}
where $\psi_i$ and $\bar{\psi}_i$ are the one-electron and its Dirac conjugated wave functions, respectively. The initial and final states of bound electrons are denoted by the indeces $i$ and $f$, $\gamma^\mu$ are the Dirac matrices and $x_i=(t_i,\vec{r}_i)$ represents the four-space coordinate vector for each electron. The function $D_{\mu\nu}^\beta(x_1,x_2)$ is the thermal part of photon propagator, see \cite{Dol,Don}.

\begin{figure}[hbtp]
	\centering
	\includegraphics[scale=0.15]{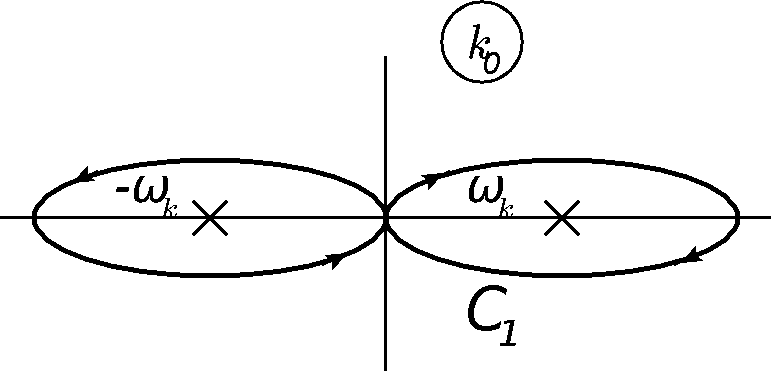} 
	\caption{Integration contour $C_1$ in $k_0$ plane of Eq. (\ref{2}).Arrows on the contour define the pole-bypass rule. The poles $\pm\omega_k$ are denoted with $\times$ marks.}
	\label{Fig-2}
\end{figure}

In \cite{S-TQED} is was established that the thermal part of photon propagator $D_{\mu\nu}^\beta(x_1,x_2)$ is given by the Hadamard propagation function \cite{Akhiezer,Greiner} and, therefore, admits a different (equivalent) form:
\begin{eqnarray}
\label{2}
i D^{\mu \nu}_{\beta}(x, x') =
- 4\pi i g^{\mu\nu}\int\limits_{C_1}\frac{d^4k}{(2\pi)^4} \frac{e^{ik(x-x')}}{k^2}n_\beta(|\vec{k}|),
\end{eqnarray}
where $g^{\mu\, \nu}$ is the metric tensor, $k^2=k_0^2-\vec{k}^2$ and $n_{\beta}$ is the Planck's distribution function. The contour of integration in $k_0$-plane for Eq. (\ref{2}) is given in Fig.~\ref{Fig-2}.
Thermal photon propagator in the form Eq. (\ref{2}) has the advantage of allowing the introduction of gauges in complete analogy with the 'ordinary' QED theory, see \cite{S-TQED}. Then in the Coulomb gauge the function $D_{\mu\nu}^\beta(x_1,x_2)$ recasts into
\begin{eqnarray}
\label{3}
D^{00}_{\beta}(x, x') &=&  4\pi i \int\limits_{C_1}\frac{d^4k}{(2\pi)^4}\frac{e^{i k (x-x')}}{\vec{k}^2}n_\beta(\omega),\qquad
\\
\nonumber
D^{ij}_{\beta}(x, x') &=& 4\pi i \int\limits_{C_1}\frac{d^4k}{(2\pi)^4}\frac{e^{i k (x-x')}}{k^2}n_\beta(\omega)\left(\delta^{ij}-\frac{k^i k^j}{\vec{k}^2}\right).
\end{eqnarray}
In following calculations we restrict ourselves by the consideration of $D^{00}_{\beta}(x, x')$ only, reproducing the thermal Coulomb interaction of two charges in the way similar to the ordinary Coulomb interaction \cite{LabKlim}.

Substitution of the Coulomb part $D^{00}_{\beta}(x, x')$ into Eq. (\ref{1}), after the integration in $k_0$-plane and time variables $t_{1,2}$, yields
\begin{eqnarray}
\label{4}
S_{fi} = -2\pi i\delta\left(E_f-E_i+E_{f'}-E_{i'}\right)\frac{4e^2}{\pi}\times\qquad
\\
\nonumber
\int d^3r_1 d^3r_2\left(\bar{\psi}_f\psi_i\right)(\vec{r}_1)\int\limits_0^\infty d\kappa\,n_\beta(\kappa)\frac{\sin\kappa r_{12}}{\kappa r_{12}}\left(\bar{\psi}_{f'}\psi_{i'}\right)(\vec{r}_2).
\end{eqnarray}
Here $r_{12}\equiv |\vec{r}_1-\vec{r}_2|$ as usual. The energy shift for an arbitrary bound state arises via the relations: $\langle a'|\hat{S}|a\rangle = - 2\pi i \langle a'| U | a \rangle\delta(E_{a'}-E_a)$ and $\Delta E_a = \langle a | U | a \rangle$. Thus, one can find that the integral $\int\limits_0^\infty d\kappa\,n_\beta(\kappa)\frac{\sin\kappa r_{12}}{\kappa r_{12}}$ represents the desired thermal interaction potential averaged on the one-electron wave functions. Evaluation of the diagram in Fig.~\ref{Fig1} in case of zero temperature does not contain the Planck's distribution function and is twice less than the expression (\ref{4}). Then the integration over $\kappa$ would lead to $\pi/2r_{12}$, i.e. the interelectron Coulomb interaction.

In particular, Eq. (\ref{4}) implies a divergence of the type $\int\limits _ 0^\infty d\kappa\,n_\beta(\kappa)$ arising from a series expansion in powers of $\kappa r_{12}$. This divergence however does not depend on the state and can simply be excluded from consideration, since it vanishes in the energy difference (transition frequency). For the conciseness we omit the discussion of possible ways to regularize this divergence and employ the procedure proposed in \cite{S-TQED}, where the appropriate analytical calculations of the integral over $\kappa$ can be found. The final result was obtained as
\begin{eqnarray}
\label{5}
V^\beta(r)=-\frac{4e^2}{\pi}
\left(-\frac{\gamma}{\beta}+\frac{i }{2 r}\ln \left[\frac{\Gamma \left(1+\frac{i r}{\beta}\right)}{\Gamma \left(1-\frac{i r}{\beta}\right)}\right]\right),
\end{eqnarray}
where $\beta\equiv 1/(k_B T)$ ($k_B$ is the Boltzmann constant and $T$ is the temperature in kelvin) and $r$ represents the modulus (length) of corresponding radius vector for the electron-nucleus or electron-electron interactions. $\Gamma$ is the Gamma function and $\gamma$ is the Euler-Mascheroni constant, $\gamma\simeq 0.577216$.

In the nonrelativistic limit at low temperatures, the potential (\ref{5}) can be expanded in powers of $r\rightarrow 0$ or $\beta\rightarrow\infty$ that in the lowest order gives
\begin{eqnarray}
\label{6}
V^\beta(r) \approx -\frac{4e^2}{\pi} \left[-\frac{r^2 \zeta(3)}{3 \beta^3}+\frac{r^4 \zeta(5)}{5 \beta^5}+\dots\right],
\end{eqnarray}
where $\zeta(n)$ is the Riemann zeta function. It should be emphasized here that the expression (\ref{6}) matches exactly to the series expansion in the formula (\ref{4}) as if the divergent terms were excluded. Then, averaging the expression (\ref{6}) on two-electron wave-functions, the lower-order thermal correction can be written as
\begin{eqnarray}
\label{7}
\Delta E^\beta_A = \frac{4\zeta(3)}{3\pi\beta^3}\langle A|e^2 r_{12}^2 -Ze^2 r_1^2 - Ze^2 r_2^2|A\rangle.
\end{eqnarray}
Here $A$ represents a two-electron bound state, $r_i$ is the length of the radius vector for each bound electron, respectively, and $r_{12}$ denotes the interelectron distance. The derivation of the thermal interaction of electrons with nucleus repeats the above calculations and we omit it for the short, see \cite{S-TQED} for more details. 

The parametric estimate of $\Delta E^\beta_A$ can be obtained in the ordinary manner, it is 
 $\Delta E_a^{\beta}\sim Z^5\alpha^3/\beta^3$ for the thermal electron-nucleus interaction and $\Delta E_a^{\beta}\sim Z^4\alpha^3/\beta^3$ for the thermal electron-electron interaction in atomic units. The temperature factor is $\beta^{-1}=k_B T\sim 10^{-3}$ at room temperature and, therefore, this correction can be compared with QED corrections of the order of $\alpha^7m$ \cite{PPY-2017}. In addition, the correction (\ref{7}) should be compared with the BBR-induced Stark shift. According to the results of \cite{SLP-QED} the parametric estimate in $\alpha$ for the Stark shift is the same, but should have an extra factor of $\beta$ at low temperatures. Thus, one can expect that the values of Eq. (\ref{3}) will exceed the corresponding BBR-induced Stark shift.

For the numerical calculations in the two-electron atom, we use trial wave functions with quasirandom nonlinear parameters developed in \cite{Korobov_1999,Korobov_2000}. As a first step in testing the methods of calculation, the nonrelativistic energies of helium states were evaluated, see Table~\ref{tab:1}, which are in good agreement with \cite{Drakebook,Frolov-1998}. 
\begin{table}[ht]
\begin{center}
\caption{Nonrelativistic energies of helium states obtained in the present work in a.u. by variational method \cite{Korobov_2000}.}
\label{tab:1}
\begin{tabular}{ c  c  c }
\hline
\hline
State & Value obtained in this work & Drake \cite{Drakebook} \\
\hline
\noalign{\smallskip}
$ 1^1S $ & $ -2.9037243770 $  & $ -2.9037243770341195 $ \\
\noalign{\smallskip}
$ 2^1S $ & $ -2.1459740460 $  & $ -2.145974046054419 $ \\
$ 2^3S $ & $ -2.1752293782 $  & $ -2.17522937823679130 $ \\
\noalign{\smallskip}
$ 2^1P $ & $ -2.1238430864 $  & $ -2.123843086498093 $ \\
$ 2^3P $ & $ -2.1331641908 $  & $ -2.133164190779273 $ \\
\noalign{\smallskip}
$ 3^1S $ & $ -2.0612719897 $  & $ -2.061271989740911 $ \\
$ 3^3S $ & $ -2.0686890674 $  & $ -2.06868906747245719 $ \\
\noalign{\smallskip}
$ 3^1P $ & $ -2.0551463620 $  & $ -2.05514636209194 $ \\
$ 3^3P $ & $ -2.0580810842 $  & $  -2.05808108427428 $ \\
\noalign{\smallskip}
$ 3^1D $ & $ -2.0556207328 $  & $ -2.055620732852246 $ \\
$ 3^3D $ & $ -2.0556363094 $  & $ -2.055636309453261 $ \\
\noalign{\smallskip}
$ 4^1S $ & $ -2.0335867167 $ & $ -2.03358671703072$ \\
$ 5^1S $ & $ -2.0211767759 $ & $ -2.021176851574363$ \\
\hline
\hline
\end{tabular}
\end{center}
\end{table}

The calculated expectation values of the $r_{12}^2$, $r_1^2$ operators and the corresponding energy shift are collected in Table~\ref{tab:2}.
\begin{table}[ht]
\begin{center}
\caption{Expectation values of $r_1^2$, $r_{12}^2$ operators (in a.u.) and the corresponding energy shift Eq. (\ref{7}) at room temperature ($T=300$K) in Hz for the He($M = \infty$) atom.
}
\label{tab:2}
\begin{tabular}{ c  c  c  c  c}
\hline
\hline
\noalign{\smallskip}
State & $r_{12}^2$ in a.u. & $r_1^2$ in a.u. & $\Delta E^\beta_A$ in a.u.  & $\Delta E^\beta_A$ in Hz \\
\hline
\noalign{\smallskip}
$ 1^1S $ & $ 2.516439313 $  & $ 1.193482995 $ & $ -3.83766\cdot 10^{-16} $ & $ -2.52505 $\\
\noalign{\smallskip}
$ 2^1S $ & $ 32.30238038 $  & $ 16.08923325 $ & $ -5.44916\cdot 10^{-15} $ & $ -35.8537 $\\
$ 2^3S $ & $ 23.04619748 $  & $ 11.46432162 $ & $ -3.8778\cdot 10^{-15} $ & $ -25.5147 $\\
\noalign{\smallskip}
$ 2^1P $ & $ 31.59851603 $  & $ 15.76565497 $ & $ -5.34878\cdot 10^{-15} $ & $ -35.1933 $\\
$ 2^3P $ & $ 26.64279322 $  & $ 13.21174046 $ & $ -4.45461\cdot 10^{-15} $ & $ -29.3099 $\\
\noalign{\smallskip}
$ 3^1S $ & $ 171.8387553 $  & $ 85.89015822 $ & $ -2.91921\cdot 10^{-14} $ & $ -192.075 $\\
$ 3^3S $ & $ 137.4750832 $  & $ 68.70840131 $ & $ -2.33505\cdot 10^{-14} $ & $ -153.638 $\\
\noalign{\smallskip}
$ 3^1P $ & $ 183.7866266 $  & $ 91.87290715 $ & $ -3.12292\cdot 10^{-14} $ & $ -205.478 $\\
$ 3^3P $ & $ 164.3028806 $  & $ 82.10989293 $ & $ -2.79026\cdot 10^{-14} $ & $ -183.59 $\\
\noalign{\smallskip}
$ 3^1D $ & $ 126.4161413 $  & $ 63.17681865 $ & $ -2.1469\cdot 10^{-14} $ & $ -141.259 $\\
$ 3^3D $ & $ 126.2834766 $  & $ 63.11075331 $ & $ -2.14467\cdot 10^{-14} $ & $ -141.112 $\\
\noalign{\smallskip}
$ 4^1S $ & $ 562.8623853 $ & $ 281.4144002 $ & $ -9.56731\cdot 10^{-14} $ & $ -629.499 $\\
$ 5^1S $ & $ 1406.742766 $ & $ 703.3605371 $ & $ -2.39134\cdot 10^{-13} $ & $ -1573.42 $\\
\hline
\hline
\end{tabular}
\end{center}
\end{table}
In particular, from Table~\ref{tab:2} it follows that the thermal correction $\Delta E^{\beta}_A$, Eq. (\ref{7}), is one order of magnitude larger than the corresponding Stark shift for lowest states. However, the correction Eq. (\ref{7}) becomes the same order as the Stark shift, see \cite{farley}, for the state $ 3^1S $ and higher but it has the opposite sign.

To demonstrate the correct reasoning on the temperature corrections, the numerical data for the second term in Eq. (\ref{6}) are given in Table~\ref{tab:3}. The order of magnitude was estimated as $Z^7\alpha^5$ for the thermal interaction of electrons with nucleus interaction and, consequently, as $Z^6\alpha^5$ for the thermal electron-electron interaction in atomic units.
\begin{table}[ht]
\begin{center}
\caption{Expectation values of second term in Eq. (\ref{6}) and corresponding  energy shifts at room temperature ($T=300$K) for helium.
}
\label{tab:3}
\begin{tabular}{ c  c  c  c }
\hline
\hline
State & $r_{12}^4$ in a.u. & $r_1^4$ in a.u. & $\Delta E^{\beta(r^4)}_A$ in Hz \\
\hline
\noalign{\smallskip}
$ 1^1S $ & $ 12.98127136 $ & $ 3.973564932 $  & $ -8.10541\cdot 10^{-11} $\\
\noalign{\smallskip}
$ 2^1S $ & $ 0.1737415294\cdot 10^4 $ & $ 0.8257531786\cdot 10^3 $  & $ -4.35629\cdot 10^{-8} $\\
$ 2^3S $ & $ 0.9163894338\cdot 10^3 $ & $ 0.4284022731\cdot 10^3 $  & $ -2.21827\cdot 10^{-8} $\\
\noalign{\smallskip}
$ 2^1P $ & $ 0.1839207883\cdot 10^4 $ & $ 0.8786029923\cdot 10^3 $  & $ -4.66127\cdot 10^{-8} $\\
$ 2^3P $ & $ 0.1349634573\cdot 10^4 $ & $ 0.6388893042\cdot 10^3 $  & $ -6.58923\cdot 10^{-9} $\\
\noalign{\smallskip}
$ 3^1S $ & $ 0.4271527957\cdot 10^5 $ & $ 0.2113668547\cdot 10^5 $  & $ -1.16396\cdot 10^{-6} $\\
$ 3^3S $ & $ 0.2772173839\cdot 10^5 $ & $ 0.1368384508\cdot 10^5 $  & $ -7.51657\cdot 10^{-7} $\\
\noalign{\smallskip}
$ 3^1P $ & $ 0.5018603163\cdot 10^5 $ & $ 0.2485719672\cdot 10^5 $  & $ -1.37018\cdot 10^{-6} $\\
$ 3^3P $ & $ 0.4055841235\cdot 10^5 $ & $ 0.2006781742\cdot 10^5 $  & $ -1.10501\cdot 10^{-6} $\\
\noalign{\smallskip}
$ 3^1D $ & $ 0.2570029451\cdot 10^5 $ & $ 0.1268705703\cdot 10^5 $  & $ -6.96961\cdot 10^{-7} $\\
$ 3^3D $ & $ 0.2565480602\cdot 10^5 $ & $ 0.1266449525\cdot 10^5 $  & $ -6.95715\cdot 10^{-7} $\\
\noalign{\smallskip}
$ 4^1S $ & $ 0.4323006495\cdot 10^6 $ & $ 0.2154350194\cdot 10^6 $ & $ -1.19492\cdot 10^{-5} $\\
$ 5^1S $ & $ 0.2624327529\cdot 10^7 $ & $ 0.1310386371\cdot 10^7 $ & $ -7.28243\cdot 10^{-5} $ \\
\hline
\hline
\end{tabular}
\end{center}
\end{table}
Thus, it can be found that the thermal correction corresponding to the second term in Eq. (\ref{6}) is negligible.

As the next step in the analysis of the diagram in Fig.~\ref{Fig1}, the thermal correction on finite size of nucleus (NS) can be obtained. This correction occurs by the introducing the charge distribution of the nucleus $\rho\approx 1 - \frac{\vec{k}2}{6}r_N^2$ in momentum space, where $r_N^2$ denotes the mean square value of the nuclear charge radius. Derivation of thermal correction on the finite size of nucleus can be found in \cite{S-TQED}, where the lowest order effect was found in the form:
\begin{eqnarray}
\label{8}
\Delta E^{\beta(NS)}_A = \frac{32 Ze^2}{15}\frac{\zeta(5)}{\beta^5}r_N^7\left|\psi_A(0)\right|^2.
\end{eqnarray}

To evaluate this correction, one can separate out the ordinary (non-thermal) NS correction of lowest order:
\begin{eqnarray}
\label{9}
\Delta E^{\beta(NS)}_A =C_\beta \frac{2\pi}{3}Z e^2\left|\psi_A(0)\right|^2r_N^2,
\\
\nonumber
C_\beta= \frac{16}{5\pi}\zeta(5)\frac{r_N^5}{\beta^5}.
\end{eqnarray}
The dimensionless coefficent $C_\beta$ can be estimated via the Compton wavelength $\lambdabar = \hbar/m_e c$ as follows
\begin{eqnarray}
\label{10}
C_\beta \equiv \frac{16}{5\pi}\zeta(5)\frac{r_N^5}{\lambdabar^5\beta^5}.
\end{eqnarray}
where $r_N$ is given in fm and $\beta = m_e c^2/k_B T$ (in SI units) can be translated in appropriate units. The thermal correction on the finite size of nucleus, Eq. (\ref{8}), is insignificant at relevant temperatures, since the coefficient $C_\beta$ is equal to $4.0765\cdot 10^{-50} r_N^5 $ $ \rm{fm}^{-5} $ at room temperature.

The result Eq. (\ref{7}) can be extended to a negative hydrogen atom. The corresponding expectation values of the operators are $r_1^2=11.91369967805$ and $r_{12}=25.2020252912$ in atomic units for the ground state, respectively, see \cite{Drakebook}. Then, the thermal correction (\ref{7}) gives the energy shift $\Delta E^\beta_A = 1.53755$ Hz at the temperature $300$ K. In contrast to the helium atom this correction has the opposite sign. Since the hydrogen anion is of particular interest in studying the atmosphere of stars, we also give the value $\Delta E^\beta_A = 12300.4$ Hz at an effective temperature of photosphere $6000$ K, and the same correction is equal to $1.92194\cdot 10^{11}$ Hz (or $0.00079485$ eV) at the corona temperature ($1.5\cdot 10^6$ K) of the Sun.


The results of this paper were obtained within the $S$-matrix formalism. It is found that the replacement of the ordinary photon line by the thermal one gives rise to the thermal potential. Since the thermal potential Eq. (\ref{5}) was never recognized before, the new type of thermal corrections is introduced here. The theory presented in this paper for the thermal corrections of lowest order reduces to evaluation of one-particle operators to thermal electron-electron and electron-nuclear interactions in atomic systems with few electrons. Thus, the corresponding mathematical derivations repeat the case of one-electron atom and, as a consequence, the result Eq. (\ref{6}) and thermal correction of lowest order Eq. (\ref{7}) were obtained within the nonrelativistic limit and the point-nucleus assumption. Going beyond the approximation of a point nucleus the NS thermal correction can be considered, see Eq. (\ref{8}), which turns out to be negligibly small.

The parametric estimation of thermal correction Eq. (\ref{6}) is the same as for the well-known Stark shift induced by the BBR field. However, this correction differs by the factor of temperature. The numerical results are given in Table~\ref{tab:2} for the different bound states of helium atom. The values in Tables~\ref{tab:2} and \ref{tab:3} establish that the thermal corrections arising from the thermal photon exchange of electron with the nucleus and other electrons reach a level of few kHz for low-lying states in ${}^4$He atom at the room temperature. The contribution grows with an increase of the principal quantum number of the bound state repeating the behaviour in the one-electron atomic system, see \cite{S-TQED}. We limited our calculations upto the $5^1S$ state, since the descipancy between the numerical values of the bound energies (comparing with the results of \cite{Drakebook}) becomes visible, see Table~\ref{tab:1}. Although such accuracy is not required (ten digits after the decimal point), the calculations, in principle, can be continued to higher excited states. We leave this problem for future works.

It should noted here that the evaluation of diagrams with two, three and etc thermal photon exchange between electrons or an electron and a nucleus would lead to a thermal correction of the same order in powers of $\alpha$ according to the QED theory. However, each thermal photon line would produce an additional factor of temperature (cubic in our case) and, therefore, such diagrams should lead to thermal corrections of the next orders of smallness. Partially, a rapid decrease in temperature is confirmed by an estimate of the thermal correction corresponding to the second term in the formula (\ref{6}). The numerical results for this correction are compiled in Table~\ref{tab:3} which shows the insignificance of the effect.

Finally, the data listed in Tables~\ref{tab:2}, \ref{tab:3} were obtained for a temperature $300$ K. They can be easily extended to other temperatures by dint of the factors $(\frac{T}{300})^3$ and $(\frac{T}{300})^5$, respectively. For example, at temperature $77$ K (the boiling point of nitrogen) the coefficient $(\frac{T}{300})^3$ is $0.0169086$. Thus, one can expect that the presence of corrections resulting from the thermal photon exchange between bound electrons as well as bound electron and nucleus can be verified experimentally in a helium atom by varying the temperature. In conclusion, this new type of thermal corrections can serve to test fundamental interactions on helium and helium-like ions.

\section*{Acknowledgements}
This work was supported by Russian Foundation for Basic Research (grant 20-02-00111). T. Z. acknowledges foundation for the advancement of theoretical physics and mathematics "BASIS". The authors are indebted to V. I. Korobov for permission to use the {\it Fortran} code for the construction of the He variational wave functions.

\bibliography{mybibfile}

\end{document}